\documentclass[reprint,superscriptaddress, amsmath,amssymb,pra, aps,floatfix,longbibliography,nofootinbib]{revtex4-1}
\usepackage[utf8]{inputenc}

\usepackage{hyperref}
\usepackage{dcolumn}
\usepackage{bm}
\usepackage{hyperref}
\usepackage[dvipsnames]{xcolor}
\usepackage{graphicx}
\graphicspath{ {./images/} }
\usepackage[capitalise]{cleveref}
\usepackage{textcomp}
\usepackage{pdfcomment}
\usepackage{bold-extra}
\usepackage{upgreek}

\DeclareUnicodeCharacter{2009}{\,}
\hypersetup{pdftoolbar=true,        
 pdfmenubar=true,        
 pdffitwindow=false,     
 pdfstartview={fitH},    
 pdftitle={},    
 pdfauthor={Daniel Pimbi},     
 pdfsubject={DWsynthesis},   
 pdfcreator={Gregory Moille},   
 pdfproducer={LaTeX}, 
 pdfkeywords={Version 1.0},
 colorlinks=true,       
 linkcolor=blue,          
 citecolor=blue,        
 filecolor=blue,      
 urlcolor=blue}

\usepackage{threeparttable}

\makeatletter
\renewcommand{\@biblabel}[1]{#1. }
\renewcommand{\@dotsep}{500}
\renewcommand{\@pnumwidth}{0em}
\renewcommand{\l@figure}[2]{
\@dottedtocline{1}{1.5em}{2em}{Figure #1}{}\vspace{15pt}}

\usepackage{tabularx}
\usepackage{makecell}

\makeatletter
\def\section{%
  \@startsection{section}{1}{\z@}%
  {-3.5ex \@plus -1ex \@minus -.2ex}%
  {2.3ex \@plus .2ex}%
  {\normalfont\large\bfseries\raggedright}%
}

\def\@hangfrom@section#1#2#3{\@hangfrom{#1#2}#3}
\def\@hangfroms@section#1#2{#1#2}
\makeatother

\usepackage[normalem]{ulem}

\begin{document}

\title{High-power visible laser via injection locking to a photonic integrated circuit optical parametric oscillator}

\author{Daniel Pimbi} 
\affiliation{Microsystems and Nanotechnology Division, Physical Measurement Laboratory, National Institute of Standards and Technology, Gaithersburg, Maryland 20899, USA}
\affiliation{Joint Quantum Institute, NIST/University of Maryland, College Park, Maryland 20742, USA}

\author{Zhiquan Yuan}   
\affiliation{Microsystems and Nanotechnology Division, Physical Measurement Laboratory, National Institute of Standards and Technology, Gaithersburg, Maryland 20899, USA}
\affiliation{Joint Quantum Institute, NIST/University of Maryland, College Park, Maryland 20742, USA}
\affiliation{Department of Electronic Engineering, Tsinghua University, Beijing, China, 100084}

\author{Ashish Chanana}   
\affiliation{Microsystems and Nanotechnology Division, Physical Measurement Laboratory, National Institute of Standards and Technology, Gaithersburg, Maryland 20899, USA}

\author{Usman A. Javid}   
\affiliation{Microsystems and Nanotechnology Division, Physical Measurement Laboratory, National Institute of Standards and Technology, Gaithersburg, Maryland 20899, USA}
\affiliation{Joint Quantum Institute, NIST/University of Maryland, College Park, Maryland 20742, USA}

\author{Xiyuan Lu}   
\affiliation{Microsystems and Nanotechnology Division, Physical Measurement Laboratory, National Institute of Standards and Technology, Gaithersburg, Maryland 20899, USA}
\affiliation{Joint Quantum Institute, NIST/University of Maryland, College Park, Maryland 20742, USA}

\author{Jordan Stone}
\affiliation{Microsystems and Nanotechnology Division, Physical Measurement Laboratory, National Institute of Standards and Technology, Gaithersburg, Maryland 20899, USA}

\author{Kartik Srinivasan} \email{kartik.srinivasan@nist.gov}
\affiliation{Microsystems and Nanotechnology Division, Physical Measurement Laboratory, National Institute of Standards and Technology, Gaithersburg, Maryland 20899, USA}
\affiliation{Joint Quantum Institute, NIST/University of Maryland, College Park, Maryland 20742, USA}

\date{\today}

\begin{abstract}
\noindent Kerr photonic integrated circuit optical parametric oscillators (PIC~OPOs) provide a promising platform for coherent visible-light generation, offering broad wavelength accessibility and low frequency noise for applications in quantum information processing, precision metrology, spectroscopy, and sensing. 
However, achieving large fiber-coupled output powers and high conversion efficiency remains challenging, impacting the practical deployment of these sources. Here we demonstrate optical injection locking of a commercially available visible Fabry-Pérot (FP) laser diode to a PIC~OPO, realizing a hybrid laser that combines the complementary strengths of nonlinear frequency conversion with those of semiconductor laser technologies. Using 25~$\mu$W of injected PIC~OPO signal, we generate coherent visible-light at 635~nm with a fiber-coupled output power of $\approx$~40~mW and side-mode suppression ratio $\gtrsim$~36.3~dB. The injection-locked FP laser faithfully inherits the low frequency noise characteristics of the PIC~OPO signal, achieving a frequency noise floor of 714~Hz$^2$/Hz. We further demonstrate $\gtrsim1$~GHz wide locking bandwidths that enable stable injection locking without active PIC~OPO stabilization, together with coarse wavelength tunability over 6~nm while preserving both high output power and spectral purity. Our results establish optical injection locking with nonlinear light sources as a compelling option for coherent, high-power, and wavelength-agile integrated coherent visible sources.
\end{abstract}

\maketitle


\section{Introduction}

{
  \let\thefootnote\relax 
  \footnotetext{\hspace{-0.1in}$^{\ddagger}$~This document is preliminary and is intended for peer review conducted by a journal.}
}

\begin{figure*}[t]
\centering\includegraphics[width=0.98\linewidth]{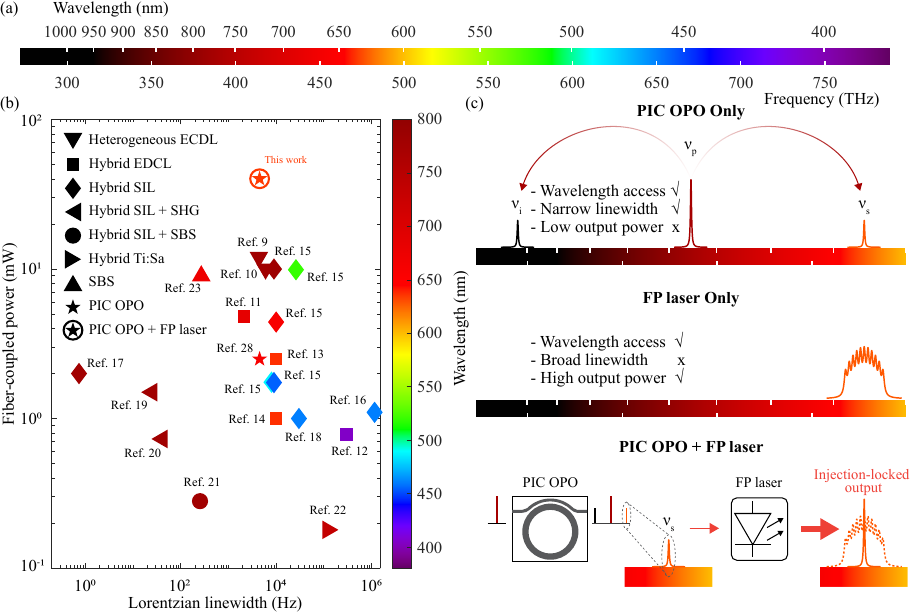}
\caption{\textbf{Hybrid PIC~OPO–FP laser based on optical injection locking for coherent high-power visible-light generation.} \textbf{(a)} Colored map of the electromagnetic spectrum spanning ultraviolet to near-infrared wavelengths. \textbf{(b)} Performance comparison of representative PIC visible laser technologies (380~nm to 785~nm) in terms of fiber-coupled output power and Lorentzian linewidth. Marker color denotes the operating wavelength, as indicated by the color scale on the right. More details are provided in the supplementary information. \textbf{(c)} Concept of OIL of an FP laser diode to a PIC~OPO signal. The top panel illustrates the spectrum of a PIC~OPO, where a pump at frequency $\nu_\text{p}$ generates coherent visible and near-infrared light at frequencies $\nu_\text{s}$ and $\nu_\text{i}$, respectively. Although PIC~OPOs provide broad wavelength accessibility and high spectral purity, they typically deliver only limited fiber-coupled output power. In contrast, FP laser diodes provide high fiber-coupled output power but generally exhibit broad linewidths (middle panel). By OIL an FP laser diode to a PIC~OPO signal, the resulting hybrid laser inherits the narrow linewidth of the PIC~OPO while retaining the high output power of the FP laser (bottom panel), thereby overcoming the limited fiber-coupled output power of PIC~OPOs.
} 
\label{Fig1}
\end{figure*}

\noindent Photonic integrated circuit (PIC) coherent visible-light laser sources are becoming key enabling technologies for a broad range of applications, including quantum computing \cite{niffenegger_integrated_2020}, quantum networks \cite{zheng_large_scale_2026}, optical atomic clocks \cite{newman_architecture_2019}, precision spectroscopy \cite{long_subdoppler_2024}, magnetometry \cite{pintus-integrated_2025}, biosensing \cite{liu_triplex_2018}, and microscopy \cite{tinguely_silicon_nitride_2017}. 
Several approaches have been developed to realize PIC coherent light sources across the visible and near-infrared spectral regions (Fig.~\ref{Fig1}(a))~\cite{lu_emerging_integrated_laser_2024}. Figure~\ref{Fig1}(b) focuses on representative visible PIC laser technologies spanning the 380~nm to 785~nm wavelength range in terms of fiber-coupled output power and Lorentzian linewidth \cite{zhang_photonic_2023, castro_integrated_2025, franken_hybrid_integrated_2021, franken_widely_tunable_2025, winkler_widely_2024, schrinner_hybrid_external_2024, corato_widely_2023, siddharth_near_ultraviolet_2022, isichenko_sub_hz_fundamental_2024, siddharth_narrow_2025, li_high_coherence_2023, clementi_a_chip_scale_2023, luo_visible_brillouin_2025, wang_photonc_2023,  chauhan_visible_2021}. The most direct approach is to integrate optical gain media onto a PIC, either through heterogeneous integration to realize external-cavity diode laser (ECDL) configurations~\cite{zhang_photonic_2023, castro_integrated_2025}, which combines multiple material platforms during wafer-level fabrication, or through hybrid integration \cite{franken_hybrid_integrated_2021, franken_widely_tunable_2025, winkler_widely_2024, schrinner_hybrid_external_2024}, where prefabricated gain or laser chips are integrated with passive photonic circuits during device assembly. These approaches have enabled integrated lasers that use a variety of gain media, including III–V compound semiconductors and, more recently, crystalline gain media such as Ti:sapphire \cite{wang_photonc_2023, yang_titanium_2024}. In parallel, techniques such as self-injection locking (SIL) \cite{corato_widely_2023, siddharth_near_ultraviolet_2022, isichenko_sub_hz_fundamental_2024, li_high_coherence_2023, clementi_a_chip_scale_2023, siddharth_narrow_2025} and stimulated Brillouin scattering (SBS) \cite{chauhan_visible_2021} have further reduced frequency noise, enabling exceptionally high spectral purity. Another approach for realizing low-noise visible PIC lasers is through nonlinear optical frequency conversion. Among the various possible processes, Kerr optical parametric oscillation (OPO) is particularly flexible with respect to wavelength accessibility, with the potential to generate a broad range of visible and short near-infrared wavelengths from a near-infrared pump through control of phase- and frequency-matching conditions (top panel of Fig.~\ref{Fig1}(c))~\cite{lu_opo_review_2025}. 
In addition to their broad wavelength accessibility, photonic integrated circuit optical parametric oscillators (PIC~OPOs) generate highly coherent visible-light with Lorentzian linewidths that can be in the $\approx$~1~kHz range \cite{black_optical_parametric_2022}, depending on the pump properties. Despite these advances, achieving high fiber-coupled output power at visible wavelengths is a major challenge \cite{sun_parasitic_2026}. Although on-chip output powers up to 5~mW and conversion efficiencies approaching $\approx 15$\% can be achieved through carefully engineered resonator–waveguide coupling, further increasing the output power is often limited by competing nonlinear processes at high pump powers~\cite{stone_efficient_opo_2022}.
A recent demonstration focused on overcoming those processes and reported on-chip visible OPO power near 10~mW, but chip-to-fiber coupling losses limited the fiber-coupled output power to $\approx$~5 mW \cite{sun_efficient_hopo_2025}, restricting the practical deployment of visible PIC OPOs in applications that simultaneously require broad wavelength accessibility, high spectral purity, and high fiber-coupled output power \cite{lu_emerging_integrated_laser_2024}.

\begin{figure*}[ht]
\centering\includegraphics[width=0.97\linewidth]{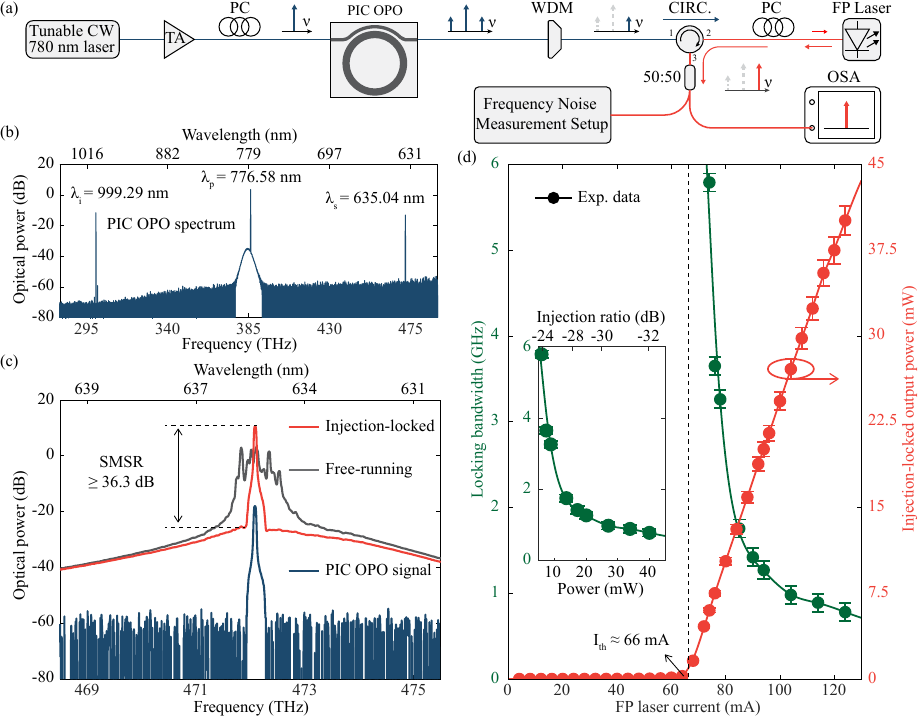}
\caption{\textbf{Experimental demonstration of OIL of an FP laser diode to a PIC~OPO  signal.} \textbf{(a)} Experimental setup. CW, continuous-wave laser; TA, tapered amplifier; PC, polarization controller; WDM, wavelength-division multiplexer; CIRC., optical circulator; OSA, optical spectrum analyzer. \textbf{(b)} Measured PIC~OPO spectrum. \textbf{(c)} Optical spectra of the injection-locked FP laser (red), the free-running FP laser (gray), and the injected PIC~OPO signal (blue). The blue trace is a zoomed-in view of the PIC~OPO signal spectrum shown in panel (b). The injection-locked FP laser exhibits an SMSR $\gtrsim$~36.3~dB while delivering a fiber-coupled output power of $\approx$~40~mW using 25~$\mu$W of injected PIC~OPO signal power. \textbf{(d)} Measured fiber-coupled output power of the injection-locked FP laser (red, right axis) and the corresponding locking bandwidth (green, left axis) as a function of FP laser drive current. The inset shows the locking bandwidth as a function of fiber-coupled output power (bottom axis) and injection ratio (top axis). Error bars for the output power represent the absolute measurement uncertainty of the power detector, whereas error bars for the locking bandwidth denote one standard deviation values obtained from frequency measurements and are primarily limited by the wavemeter accuracy. Throughout this manuscript, optical spectra plotted in dB are referenced to 1~mW (i.e., dBm).
} 
\label{Fig2}
\end{figure*}

Whereas PIC~OPOs excel in broad wavelength accessibility and high spectral purity, Fabry–Pérot (FP) laser diodes operating in the visible are commercially available, cost-effective, and capable of delivering tens of milliwatts of fiber-coupled output power. However, their free-running emission typically consists of multiple longitudinal modes (middle panel of Fig.~\ref{Fig1}(c)) and exhibits large frequency noise. Thus, although FP laser diodes provide high fiber-coupled output power, their poor spectral purity limits their use in applications demanding narrow-linewidth visible-light. The complementary strengths of PIC OPOs and FP laser diodes make them natural candidates for optical injection locking (OIL). In OIL, a low-noise seed laser is injected into a high-power laser, forcing the injection-locked laser to oscillate at the injected signal's frequency while retaining its intrinsic output power \cite{liu_optical_injection_2020}. OIL has been extensively investigated in semiconductor, fiber, and solid-state laser systems for linewidth reduction, power scaling, and frequency stabilization \cite{liu_optical_injection_2020, liu_recent_advances_2023, lau_strong_optical_2008}. However, its application to PIC nonlinear visible-light laser sources has yet to be demonstrated as a strategy for overcoming the limited fiber-coupled output power.

Here, we propose a hybrid approach based on OIL of an FP laser diode to a PIC~OPO signal (bottom of Fig.~\ref{Fig1}c). The resulting hybrid laser inherits the broad wavelength accessibility and high spectral purity of the PIC~OPO while retaining the high fiber-coupled output power of the FP laser, thereby combining these complementary attributes within a single visible-light laser source. We achieve coherent visible-light generation at 635~nm with a fiber-coupled output power of $\approx$~40~mW and a side-mode suppression ratio (SMSR) $\gtrsim$~36.3~dB using only 25~$\mu$W of injected seed power, corresponding to an injection ratio as low as -32~dB. We further show that the injection-locked FP laser inherits the low frequency noise characteristics of the PIC~OPO signal, exhibiting a frequency noise floor of 714~Hz$^2$/Hz. In addition, we demonstrate $\gtrsim1$~GHz locking bandwidths together with coarse wavelength tunability spanning 6~nm. These large bandwidths enable stable injection locking without active frequency stabilization of the PIC OPO, substantially simplifying implementation of the hybrid laser. More broadly, this concept is readily extendable across the visible spectral range accessible to PIC OPOs and commercially available FP laser diodes, establishing an alternative strategy for realizing coherent, high-power, and wavelength-agile visible-light laser sources for quantum technologies, precision metrology, spectroscopy, and sensing.

\section{Results}
The experimental setup is shown in Fig.~\ref{Fig2}(a). A tunable continuous-wave laser operating near 780~nm was amplified and coupled to a PIC~OPO. Figure~\ref{Fig2}(b) shows a PIC~OPO spectrum measured at the chip output, featuring a visible signal at $\lambda_\text{s}=635.04$~nm and a near-infrared idler at $\lambda_\text{i}=999.29$~nm generated from a pump at $\lambda_\text{p}=776.58$~nm. The PIC OPO used in this work is based on previously reported dispersion-engineered silicon nitride microring resonators, which provide access to a broad spectral range spanning 590~nm to 1150~nm through a combination of device geometry and pump-wavelength tuning \cite{stone_efficient_opo_2022}. A wavelength-division multiplexer was used to isolate the signal from the pump and idler, after which the signal was routed through an optical circulator to injection-lock a commercially available fiber-coupled FP laser diode operating near 635 nm. The injected PIC~OPO signal power, measured immediately before the FP laser, was 25~$\mu$W. The injection-locked output was characterized in terms of its optical spectrum, output power, and frequency noise. Figure~\ref{Fig2}(c) compares the optical spectra of the PIC OPO signal, the free-running FP laser, and the injection-locked FP laser. The FP laser diode was temperature controlled to align one of its longitudinal modes with the PIC~OPO signal wavelength. In the free-running state, the FP laser exhibits the characteristic multimode spectrum associated with broad-linewidth operation and delivers a fiber-coupled output power of $\approx$~40~mW. Upon injection locking, the multimode emission collapses into a single-frequency lasing output with an SMSR $\gtrsim$~36.3~dB, with the wavelengths at which this measurement is taken indicated in the figure. This spectral collapse suggests that the FP laser has been locked to the injected PIC~OPO signal. The injection-locked FP laser delivered almost the same fiber-coupled output power of $\approx$~40~mW as in the free-running state, corresponding to an injection ratio as low as $-32$~dB. The injection ratio is calculated as $10 \log_{10} (P_{\text{inj}} / P_{\text{FP}})$, where $P_{\text{inj}}$ and $P_{\text{FP}}$ denote the injected PIC~OPO signal power and the fiber-coupled output power of the FP laser, respectively \cite{liu_optical_injection_2020}.

Next, we investigated the dependence of the injection-locked output power and locking bandwidth on the FP laser drive current (Fig.~\ref{Fig2}(d)). As the drive current increased from the lasing threshold of 66~mA to the maximum operating current of 124~mA, the fiber-coupled output power increased monotonically, reaching 40~mW at the highest drive current. The locking bandwidth was quantified as the range of PIC~OPO signal frequencies (tuned via the pump frequency) over which injection locking occurs. 
As shown in Fig.~\ref{Fig2}(d), the locking bandwidth decreased with increasing drive current, consistent with the increasing intracavity field of the FP laser relative to the injected PIC~OPO signal at higher output powers \cite{liu_optical_injection_2020}. Nevertheless, $\gtrsim1$~GHz locking bandwidths were maintained across the entire operating range, providing sufficient tolerance to frequency detuning for stable injection locking without the need for active frequency stabilization of the PIC~OPO. The inset shows the locking bandwidth as a function of output power and injection ratio, demonstrating stable injection locking at injection ratios as low as $-32$~dB.  

\begin{figure}[t]
\centering\includegraphics[width=0.98\linewidth]{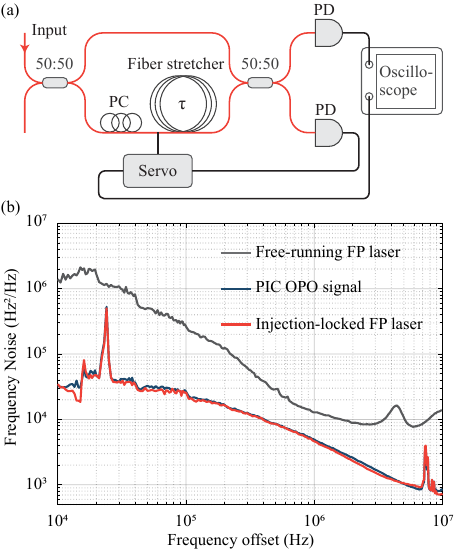}
\caption{\textbf{Frequency noise characterization.} \textbf{(a)} Experimental setup of the delayed self-homodyne measurement. PD, photodiode. \textbf{(b)} Frequency noise spectra of the free-running FP laser (gray), the PIC~OPO signal (blue), and the injection-locked FP laser output (red). Under injection locking, the FP laser inherits the low frequency noise characteristics of the PIC~OPO signal, with both reaching a frequency noise floor of $\approx$~714~$\text{Hz}^2/\text{Hz}$ at an offset frequency of 9.4~MHz.
} 
\label{Fig3}
\end{figure}

A crucial aspect of OIL is its ability to transfer the spectral purity of a low-noise PIC~OPO signal to the high-power FP laser. To quantify this coherence transfer, we measured the frequency noise of the free-running FP laser, the PIC~OPO signal, and the injection-locked FP laser using a modified delayed self-homodyne technique based on the method in Ref.~\cite{li_high_coherence_2023}, as shown in Fig.~\ref{Fig3}(a) (see Supplementary Information for details). The measured spectra are presented in Fig.~\ref{Fig3}(b). The free-running FP laser exhibits higher frequency noise across the entire measured offset frequency range. In contrast, the frequency noise spectrum of the injection-locked FP laser is nearly indistinguishable from that of the PIC~OPO signal, confirming efficient transfer of spectral purity. Both the PIC~OPO signal and the injection-locked FP laser reach a frequency noise floor of $\approx$~714~Hz$^2$/Hz at an offset frequency of 9.4~MHz, corresponding to a Lorentzian linewidth of $\approx$~4.5~kHz (see Supplementary Information for more details). Achieving this linewidth at 40~mW of output power compares favorably with other visible wavelength laser approaches in PICs as shown in Fig.~\ref{Fig1}(b).

\begin{figure}[t]
\centering
\includegraphics[width=0.98\linewidth]{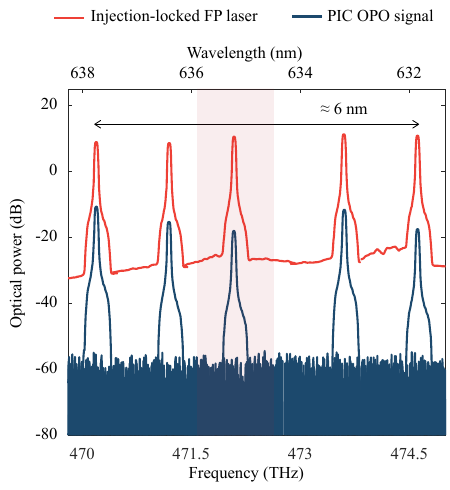}
\caption{\textbf{Coarse wavelength tuning of the injection-locked FP laser.} Overlaid optical spectra of multiple PIC~OPO operating states (blue) that were sequentially accessed to maintain stable injection locking of the FP laser (red) over a 6 nm wavelength tuning range enabled by temperature tuning of the FP laser diode. From left to right, the FP laser temperature was set to \{37.39, 36.15, 27.07, 19.26, 11.30\}$^{\circ}$C at a fixed drive current of 124~mA. The shaded spectrum reproduces the data shown in Fig.~\ref{Fig2}(c).
}
\label{fig4}
\end{figure}

In addition to high output power and low frequency noise, many practical applications require integrated visible laser sources that provide both fine and coarse wavelength tunability. The $\gtrsim1$~GHz locking bandwidths demonstrated in Fig.~\ref{Fig2}(d) enable fine frequency tuning that is essential for applications such as precision spectroscopy and addressing narrow atomic transitions. We further demonstrated coarse wavelength-tuning capability of the hybrid laser. This was achieved by accessing multiple OPO states through a combination of device geometry and pump-laser tuning and, for each OPO state, maintaining stable injection locking by adjusting the temperature of the FP laser. As shown by the overlaid optical spectra in Fig.~\ref{fig4}, coarse wavelength tuning spanning 6~nm was achieved, covering the full thermal tuning range of the FP laser. Throughout this range, the injection-locked emission remained spectrally aligned with the corresponding PIC~OPO seed while maintaining nearly the same fiber-coupled output power. 
The slight variations in the injection-locked output spectra are primarily attributed to the temperature dependence of the FP laser output power, whereas the variations in the PIC~OPO signal spectra, arise mainly from changes in the nonlinear conversion efficiency of the microresonator. These variations do not affect the operation of the hybrid laser since we remain within the range of injection ratios/locking bandwidths shown in Fig.~\ref{Fig2}(d), demonstrating the robustness of stable OIL despite changes in the PIC~OPO conversion efficiency.

\section{Discussion}
We have demonstrated OIL as an effective strategy for overcoming the limited fiber-coupled output power of PIC~OPOs operating in the visible.
OIL simultaneously addresses the principal limitations of two complementary photonic technologies by combining the spectral purity of a PIC~OPO with the high output power and technological maturity of a commercially available FP laser diode. From the PIC~OPO perspective, it relaxes demands on its output power and efficiency, enabling a designer to focus on other performance metrics such as wavelength access and tunability, or to optimize for low PIC~OPO threshold so long as the output power is sufficient to reach the modest ($\approx25$~$\mu$W) seed power we have used. In comparison to self-injection locked lasers or PIC-integrated external cavity tunable lasers, our approach directly leverages commercially available FP laser diodes without requiring optical feedback from an external cavity.

The generality of this concept is further supported by successful optical injection locking of additional commercially available visible FP laser diodes (see Supplementary Information), demonstrating its broad compatibility with existing semiconductor laser technologies. In addition, while we focus on PIC~OPOs with an output signal wavelength in the 632~nm to 642~nm range, Kerr PIC~OPOs have been demonstrated for visible wavelengths between 532~nm and 780~nm~\cite{sun_green_gap_2024}, enabling OIL with many additional commercial FP laser options. 

The concept of using OIL to boost the optical power after a nonlinear process is not limited to Kerr PIC~OPOs and is readily extendable 
to other $\chi^{(3)}$ or $\chi^{(2)}$ processes and wavelengths, providing a general strategy when laser gain media at the relevant wavelengths are available, but high power, low-noise operation is a challenge. 
Furthermore, recent advances with heterogeneous integration of III-V semiconductor gain with silicon nitride PICs across a variety of wavelengths~\cite{lu_emerging_integrated_laser_2024,tran_extending_2022, zhang_photonic_2023} suggest that our approach can be fully chip-integrated. The substantially increased output power can further enable nonlinear optical processes that operate at higher powers than that typically provided by a PIC~OPO alone, including SBS~\cite{luo_visible_brillouin_2025} for improved noise performance. Beyond these technological opportunities, further work may include a studies of the physical mechanisms governing coherence transfer in injection-locked semiconductor lasers, including frequency noise transfer and linewidth narrowing.



\medskip
\noindent\textbf{Funding} NIST work was funded solely by the United States government, in part through the NIST-on-a-chip program.

\medskip
\noindent \textbf{Acknowledgements --} The authors acknowledge David Long for helpful discussions.

\medskip
\noindent \textbf{Disclosures} The authors declare no conflicts of interest.

\medskip
\noindent \textbf{Data Availability} Data underlying the results presented in this paper may be obtained from the authors upon reasonable request.

\clearpage
\bibliographystyle{osajnl}
\bibliography{OIL_POWER_arxiv}


\clearpage 
\onecolumngrid

\setcounter{section}{0}
\setcounter{figure}{0}
\setcounter{table}{0}
\setcounter{equation}{0}

\renewcommand{\thesection}{S\arabic{section}} 
\renewcommand{\thesubsection}{S\arabic{subsection}} 
\renewcommand{\thefigure}{S\arabic{figure}} 
\renewcommand{\thetable}{S\arabic{table}} 
\renewcommand{\theequation}{S\arabic{equation}} 

\renewcommand{\figurename}{Figure}

\section*{Supplementary Information: High-power visible laser via injection locking to a photonic integrated circuit optical parametric oscillator} 

\section{Microresonator design and characterization}
\begin{figure}[ht]
\centering
\includegraphics[width=0.98\linewidth]{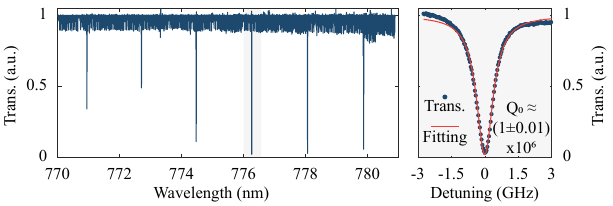}
\caption{\textbf{Measured pump-band transmission spectrum of the PIC~OPO device.} The shaded pump resonance and the corresponding nonlinear least-squares fit yield an intrinsic quality factor Q$_\text{0}$ of $\approx$~1×10$^6$ and indicate near-critical coupling. The displayed uncertainty in the figure is a one standard deviation value from repeated measurements.}
\label{S1}
\end{figure}

The photonic integrated circuit optical parametric oscillator (PIC~OPO) devices used in this work were previously developed and reported by Stone et al.~\cite{stone_efficient_opo_2022}. They consist of dispersion-engineered silicon nitride microring resonators designed for widely separated Kerr OPO between the visible and near-infrared spectral regions. Broadband pulley-waveguide couplers were employed to simultaneously achieve efficient pump coupling and high extraction efficiency for the generated signal and idler waves, enabling on-chip conversion efficiencies approaching 15~$\%$. Although the devices generated on-chip output powers of up to 5 mW, chip-to-fiber coupling losses limited the available fiber-coupled output power to less than 2.5 mW. This limitation motivated the optical injection locking (OIL) approach developed in the present work. Before performing the OIL experiments, we characterized the microresonator to verify that its performance had not degraded over time. Figure~\ref{S1} shows the normalized low-power transmission spectrum near 780 nm together with a nonlinear least-squares fit to one of the resonances. The device was designed such that optical parametric oscillation (OPO) was supported exclusively by the fundamental transverse-electric mode family. From the resonance fit, the intrinsic quality factor was extracted to be approximately $Q_\text{0}\approx1$×$10^6$, while the resonance operated close to critical coupling. This intrinsic quality factor was sufficient to sustain the high intracavity optical intensities required for an efficient Kerr OPO. The optical power budget is discussed in the following section.

\section{Optical power measurements}
\begin{figure}[ht]
\centering
\includegraphics[width=0.98\linewidth]{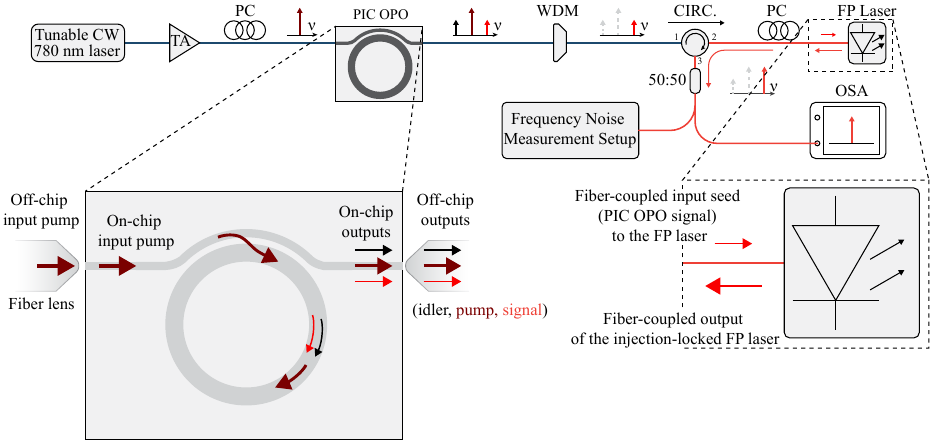}
\caption{\textbf{Experimental setup for OIL of an FP laser diode to a PIC PIC~OPO signal.} The schematic defines the optical power quantities reported throughout the manuscript. The off-chip input pump power denotes the optical power measured in the input lensed fiber immediately before coupling into the photonic chip. After accounting for the measured fiber-to-chip coupling loss, the corresponding on-chip input pump power represents the optical power coupled into the bus waveguide. Inside the microresonator, Kerr OPO generates signal and idler waves together with the residual transmitted pump. Their optical powers at the chip output are referred to as the on-chip output powers. After coupling from the chip into the output lensed fiber, the corresponding measured optical powers are referred to as the off-chip output powers. For the OIL experiments, the filtered PIC~OPO signal was directed to the FP laser through an optical circulator. The fiber-coupled PIC~OPO signal power denotes the optical power measured immediately before entering the FP laser. The injection-locked output was measured at port 3 of the optical circulator. Because this measurement includes the insertion loss between ports 2 and 3, all reported fiber-coupled output powers of the injection-locked FP laser were corrected using the independently measured circulator insertion loss so that they correspond to the optical power directly emitted from the FP laser output fiber. The same correction was applied when measuring the free-running FP laser output power.
}
\label{S2}
\end{figure}

Throughout this work, power measurements were performed using a power detector that has a specified expanded measurement uncertainty of $\pm$~3~\% (coverage factor $\kappa=2$), while optical spectra were recorded using an optical spectrum analyzer (OSA) with a resolution bandwidth of 0.05~nm. Figure~\ref{S2} summarizes the experimental setup and defines all optical power quantities reported throughout the manuscript, and Table~\ref{tab_s1} summarizes the measured and estimated powers at the various locations indicated in Fig.~\ref{S2}. The on-chip optical powers reported in Table~\ref{tab_s1} were estimated from the fiber-coupled powers using independently measured fiber-to-chip coupling losses. The measurement uncertainty is estimated to be sufficiently small that it does not affect the conclusions of this work. The error bars associated with the injection-locked output power in Fig.~\ref{Fig2}(d) were calculated using the specified $\pm$~3~\% expanded measurement uncertainty of the power detector.

\begin{table}[ht]
\centering
\begin{threeparttable}
\caption{Summary of the optical power budget for the hybrid laser.}
\label{tab_s1}
\begin{tabular}{|l|c|c|}
\hline
\textbf{Quantity}                                                   & \textbf{Wavelength (nm)} & \textbf{Power (mW)} \\ \hline
Off-chip input pump                                        & 776   & 45.7   \\ \hline
On-chip input pump                                         & 776   & 28.2   \\ \hline
On-chip output PIC~OPO signal                             & 635   & 0.227   \\ \hline
Off-chip output PIC~OPO signal                            & 635   & 0.098   \\ \hline
Fiber-coupled input seed (PIC~OPO signal) to the FP laser & 635   & 0.025   \\ \hline
Fiber-coupled output of the injection-locked FP laser      & 635   & 40.2   \\ \hline
\end{tabular}
\begin{tablenotes}[flushleft]
    \item[] \small Note: The reported power values have an expanded measurement uncertainty of $\pm$~3~\% (coverage factor $\kappa=2$), primarily due to the power detector.
\end{tablenotes}
\end{threeparttable}
\end{table}

\section{Locking bandwidth measurement}
The locking bandwidth is defined as the frequency (or equivalently wavelength) range of PIC~OPO signal frequencies over which a single longitudinal mode of the Fabry–Pérot (FP) laser remains stably injection-locked. To measure the locking bandwidth, the PIC~OPO frequency was finely tuned via the pump frequency, while its frequency was simultaneously measured using a wavemeter with a specified wavelength accuracy of $\pm$~0.0001~nm and the injection-locked optical spectrum was monitored with an OSA.
The FP laser was considered to remain injection-locked as long as its emission tracked the PIC~OPO signal continuously without mode hopping or abrupt frequency discontinuities. The locking bandwidth was determined from the total frequency excursion over which stable injection locking was maintained. The uncertainty in the measured locking bandwidth is dominated by the wavelength accuracy of the wavemeter. Accordingly, the error bars shown in Fig.~\ref{Fig2}d of the main manuscript were calculated directly from the specified $\pm$~0.0001~nm wavelength accuracy of the wavemeter.

\begin{figure}[ht]
\centering
\includegraphics[width=0.98\linewidth]{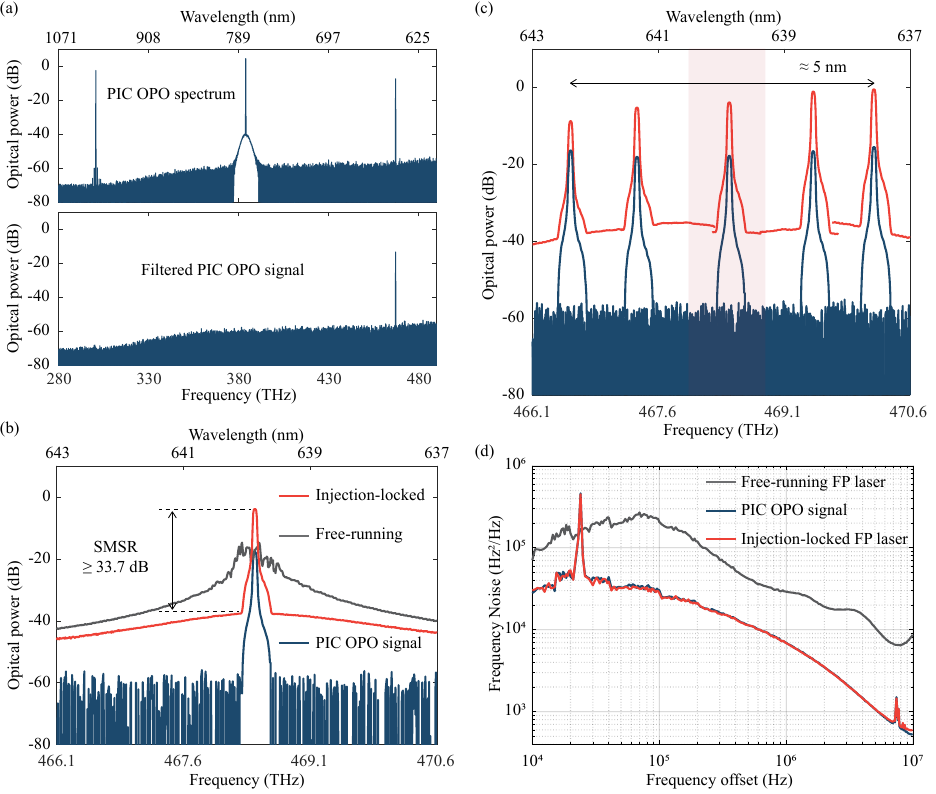}
\caption{\textbf{OIL of a commercially available FP laser diode to a PIC~OPO signal.} \textbf{(a)} Measured off-chip PIC~OPO output spectrum (top) and the filtered PIC~OPO signal used for OIL (bottom). \textbf{(b)} Optical spectra of the free-running FP laser (gray), the injected PIC~OPO signal (blue), and the injection-locked FP laser (red). The injection-locked FP laser exhibits a side-mode suppression ratio (SMSR) $\gtrsim$~33.7~dB while delivering a fiber-coupled output power of $\approx$~1~mW using 25~$\mu$W of injected PIC~OPO signal, corresponding to an injection ratio of -16~dB. A locking bandwidth of $\approx$~4~GHz was obtained, comparable to that reported for the FP laser investigated in the main manuscript. \textbf{(c)} Coarse wavelength tuning of the injection-locked FP laser over its full temperature tuning range, demonstrating approximately 5~nm of wavelength tunability. From left to right, the FP laser temperature was set to \{38.0, 32.0, 25.46, 17.40, 14.44\}$^{\circ}$C. The shaded spectra are reproduced from the data shown in (b). \textbf{(d)} Frequency noise spectra of the free-running FP laser (gray), the PIC~OPO signal (blue), and the injection-locked FP laser (red). The free-running FP laser exhibits substantially higher frequency noise across the measured offset frequency range, whereas the injection-locked FP laser faithfully inherits the low frequency noise characteristics of the PIC~OPO signal. Both the PIC~OPO signal and the injection-locked FP laser reach a frequency noise floor of approximately 537~Hz$^2$/Hz at an offset frequency of 9.6~MHz, corresponding to a Lorentzian linewidth of $\approx$ 3.4~kHz.
}
\label{S3}
\end{figure}

\section{Frequency noise measurement}
The frequency noise spectra presented in this work were measured using a modified delayed self-homodyne technique with quadrature-point locking~\cite{li_high_coherence_2023}, with the data processing performed in a manner similar to that described in ~\cite{yuan_correlated_sef-heterodyne_2022}. In this technique, the laser under test was split into two optical paths: one half of the light passed through a short reference path, while the other half propagated through a long optical-fiber delay before the two paths were recombined. The recombined signal was detected using two photodetectors. The output of one photodetector was digitized with a high-speed oscilloscope and processed digitally using Hilbert transforms and Fourier analysis to recover the single-sideband frequency noise spectrum. The second photodetector was incorporated into a feedback loop that maintained the interferometer at its quadrature operating point throughout the measurement. Assuming a Lorentzian lineshape for the short-term laser behavior, the measured frequency noise spectra were converted to the corresponding Lorentzian linewidths using the frequency noise floor according to:

\begin{equation*}
    \Delta \nu = 2 \pi S_{\text{v},}
\end{equation*}
\noindent where $S_{\text{v}}$ denotes the frequency noise floor \cite{yuan_correlated_sef-heterodyne_2022}.

\section{Optical injection locking of additional FP laser diodes}
To further demonstrate the versatility and generality of the proposed OIL architecture, OIL was performed using two additional commercially available visible FP laser diodes. Figure~\ref{S3} presents the corresponding experimental results for one of those laser diodes, whose free-running output power was $\approx$~1~mW. Similar to the FP laser diode investigated in the main manuscript, stable single-mode operation was achieved using 25~$\mu$W of injected PIC~OPO signal power. The injection-locked FP laser exhibited an SMSR $\gtrsim$~33~dB while delivering $\approx$~1~mW of fiber-coupled output power, corresponding to an injection ratio of approximately -16~dB. A locking bandwidth of $\approx$~4~GHz was measured. Frequency noise measurements further showed that the injection-locked FP laser inherited the low frequency noise characteristics of the injected PIC~OPO signal, with both exhibiting a frequency noise floor of 537~Hz$^2$/Hz at an offset frequency of 9.6~MHz, corresponding to a Lorentzian linewidth of $\approx$~3.4~kHz. A third commercially available FP laser diode, whose free-running output power was $\approx$~12~mW, was also successfully injection-locked using the same experimental approach~\cite{pimbi_oil_2026}, with output power matching the free-running power and Lorentzian linewidth of $\approx$~8.3~kHz. Collectively, these results demonstrate that the proposed OIL architecture is broadly applicable to commercially available visible FP laser diodes spanning different operating output-power levels.

\section{Comparison with previously reported PIC coherent visible-light laser sources}
Table~\ref{tab_s2} provides a comprehensive comparison of previously demonstrated PIC coherent visible-light laser sources shown in Fig.~1(b). For each work, the laser architecture, operating wavelength, Lorentzian linewidth, coarse wavelength tunability, and output power are summarized whenever available. Quantities that were not explicitly reported are indicated as "n/a." This comparison places the present work within the broader landscape of PIC coherent visible-light generation and highlights its unique combination of high fiber-coupled output power, 1~kHz~to~5~kHz Lorentzian linewidth, and nanometer-scale wavelength tunability.

\begin{table}[hb]
\centering
\caption{Characteristics of recently demonstrated PIC coherent visible-light laser sources}
\label{tab_s2}
\resizebox{\textwidth}{!}{
\begin{tabular}{|l|c|c|c|c|c|}
\hline
\textbf{Approach}      & \makecell{\textbf{Wavelength} \\ \textbf{(nm)}} & \makecell{\textbf{Lorentzian linewidth } \\  } & \makecell{\textbf{Coarse tuning}\\ \textbf{(nm)}} & \makecell{\textbf{Fiber-coupled output power} \\ \textbf{(mW)}} \\ \hline
PIC~OPO + FP laser [This work]   & 635   &  4.50~kHz  &  6  &  40.2  \\ \hline
Heterogeneous ECDL \cite{zhang_photonic_2023}   & 780   & 4.35~kHz   &  18  & 12   \\ \hline
Heterogeneous ECDL \cite{castro_integrated_2025}   & 780   & 6~kHz   &  20  & 10   \\ \hline
Hybrid ECDL \cite{franken_hybrid_integrated_2021}   & 684.4   & 2.10~kHz   &  10.8  & 4.8   \\ \hline
Hybrid ECDL \cite{franken_widely_tunable_2025}   & 405.5   & 300~kHz   &  4.4  & 0.78   \\ \hline
Hybrid ECDL \cite{winkler_widely_2024}   & 637   & 10~kHz   &  8  & 2.5   \\ \hline
Hybrid ECDL \cite{schrinner_hybrid_external_2024}   & 637   & 10~kHz   &  9  & 1   \\ \hline
Hybrid SIL \cite{corato_widely_2023}   & 785   &  9~kHz  &  12.5  &  10  \\ \hline
                                        & 660   &  10~kHz &  4.9   &  4.42  \\ \hline
                                        & 520   &  26~kHz &  6    &  9.89  \\ \hline 
                                        & 488   &  8~kHz  &  5.6  &  1.75  \\ \hline
                                        & 455   &  9~kHz  &  4.5  &  1.74 \\ \hline
Hybrid SIL \cite{siddharth_near_ultraviolet_2022}   & 461   &  1.156~MHz  &  n/a  &  1.1  \\ \hline
Hybrid SIL \cite{isichenko_sub_hz_fundamental_2024}   & 780   &  0.74~Hz  &  6  &  2  \\ \hline
Hybrid SIL \cite{siddharth_narrow_2025}   & 461   &  30~kHz  &  n/a  &  1  \\ \hline
Hybrid SIL + SHG \cite{li_high_coherence_2023}   & 780   &  25~Hz  &  n/a  &  1.5  \\ \hline
Hybrid SIL + SHG \cite{clementi_a_chip_scale_2023}   & 780   &  41~Hz  &  n/a  &  0.73  \\ \hline
Hybrid SIL + SBS \cite{luo_visible_brillouin_2025}   & 780   &  254~Hz  &  n/a  &  0.28  \\ \hline
Hybrid Ti:Sa \cite{wang_photonc_2023}   & 735   &  120~kHz  &  n/a  &  0.18  \\ \hline
SBS \cite{chauhan_visible_2021}   & 674   &  270~Hz  &  n/a  &  9  \\ \hline
\end{tabular}
}
\end{table}

\end{document}